\documentclass[twocolumn,showpacs,preprintnumbers,amsmath,amssymb]{revtex4}

\usepackage{graphicx}
\usepackage{dcolumn}
\usepackage{bm}

\begin{document}

\preprint{APS/123-QED}


\title{Absolute negative conductivity and zero-resistance states\\
in two dimensional electron systems: A plausible scenario
} 
\author{V.~Ryzhii}
\email{v-ryzhii@u-aizu.ac.jp}
 \affiliation{Computer Solid State Physics Laboratory, University of Aizu,
Aizu-Wakamatsu 965-8580, Japan}
\author{A.~Chaplik}
 \affiliation{Institute of Semiconductor Physics, RAS,
Novosibirsk 630090, Russia
}
\author{R.~Suris}
\affiliation{A.F.Ioffe Physical-Technical Institute, RAS,
St.~Petersburg 194121, Russia} 

\date{\today}

\begin{abstract}
We present a model which provides a plausible explanation
of  the effect of
zero-resistance and
zero-conductance
states in two-dimensional electron systems subjected to a magnetic field
and irradiated with microwaves observed in a number of experiments and
of  the effect main features. 
The model is based on the concept of absolute
negative conductivity associated with photon-assisted scattering of electrons
on impurities. It is shown that the main features of the effect can be attributed
to the interplay of different electron scattering mechanisms.
\end{abstract}



\maketitle

\section{Introduction}

The possibility of the states in which the dissipative electric current
in a nonequilibrium electron system (a system in which a majority of
electrons have negative effective mass)
flows in the direction opposite to the electric
field, i.e., the usual (or absolute) conductivity of the system is
negative,   was discussed by Kroemer in the late 50s~\cite{1}.
Rather realistic mechanisms of such an absolute negative conductivity (ANC)
in two- and three-dimensional   substantially nonequilibrium
electron systems (2DESs and 3DESs)
in magnetic field were considered more than three decades ago~\cite{2,3,4}.
At the same time, the mechanism of ANC in a 2DES subjected to
magnetic field and irradiated with microwaves associated with   impurity
scattering of 2D electrons accompanied by the absorption of microwave photons
was proposed by one of us~\cite{5}.
It was shown that the dissipative
conductivity is an oscillatory function of the ratio
of the microwave frequency $\Omega$ to the electron
cyclotron frequency $\Omega_c$. 
At $\Omega$ somewhat exceeding
$\Omega_c$ or a multiple of $\Omega_c$,
the photon-assisted impurity scattering of 2D electrons with
their transitions between the Landau levels (LLs) results
in a contribution to the dissipative current 
flowing opposite to
 the electric field. At sufficiently strong microwave radiation,
this scattering mechanism can dominate leading to ANC when 
$\Omega \gtrsim N\Omega_c$, where $N = 1,2,3,...$
The transformation of the dissipative current vs electric field characteristic
is schematically shown in fig.1. 
\begin{figure}
\centerline{\includegraphics[width=70mm]{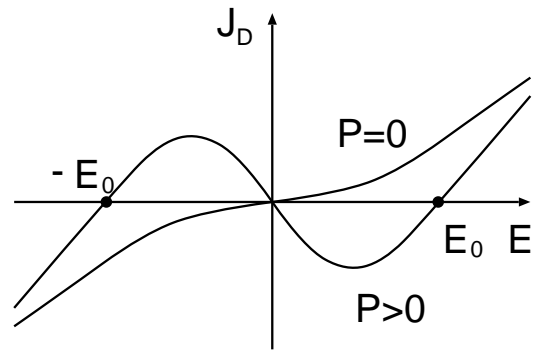}}
\caption{Schematic view of dissipative
current-voltage
characteristics $J_D = J_D(E)$ without ($P = 0$) and with ($P > 0$)
microwave irradiation.}
\end{figure}
\begin{figure}
\centerline{\includegraphics[width=80mm]{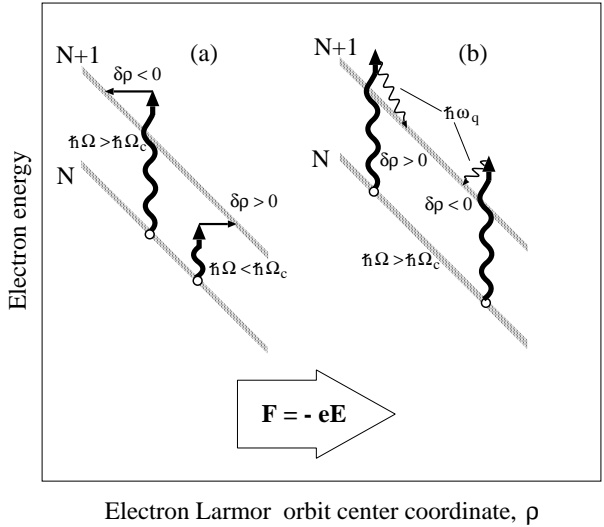}}
\caption{Inter-LL electron transitions:
(a) photon-assisted impurity for both $\Omega > \Omega_c$
and $\Omega < \Omega_c$ and (b) photon-assisted
acoustic phonon scattering mechanisms (only transitions
for $\Omega > \Omega_c$ are shown).}
\end{figure}
The effect of vanishing electrical resistance (in the Hall bar configuration)
and of vanishing electrical conductance (in the Corbino configuration)
 in a 2DES in magnetic field irradiated with microwaves
has recently been observed by Mani {\it et al.}~\cite{6}, 
Zudov {\it et al.}~\cite{7}, and Yang {\it et al.}~\cite{8}.
According to Anderson and Brinkman~\cite{9}, 
Andreev {\it et al.}~\cite{10}, and Volkov and co-workers~\cite{11}

suggested that this effect, i.e., the appearance of 
ZR-states (and ZC-states), is  attributed 
to ANC associated with photon-assisted impurity
scattering of 2D electrons (put forward in~\cite{5,12})
and an instability of homogeneous states 
in  a conductive media with ANC. The latter was noted by Zakharov~\cite{13}
and discussed in early papers on ANC in 2DESs (see, for example,~\cite{4}).
The structure of the electric-field distributions corresponding to
ZR- and ZC-states arisen as a result of the
instability is determined by the shape of the current-voltage characteristic (in particular, by the value of $E_0$) and the features of
the diffusion processes. 
Recent experimental findings~\cite{6,7,8} have stimulated a surge
of experimental~(for example,~\cite{14,15,16,17}) 
and theoretical papers (for example,~\cite{18,19,20,21,22,23,24}).
Preliminary brief overviews can be found in~\cite{25,26}.
In particular, the results of early theoretical
studies of ANC caused by photon-assisted impurity scattering
were generalized
by the inclusions of the LL broadening and
high microwave power effects~\cite{18,19,20}. A quasi-classical model
which is valid at large filling factors and sufficiently strong electric field
(or when a long-range disorder determines the dissipative current)
was developed by Vavilov and Aleiner~\cite{21}. Possible role of
photon-assisted acoustic phonon scattering was discussed in~\cite{22,23,24}. 

A theoretical model for ZR- and ZC-states should explain at least
the following details observed experimentally: (a) the phase of the magnetic-field
dependence of the resistance (dissipative conductivity), i.e.,
the positions of
maxima and minima, (b) very slow dependence of the magnitude
of the dissipative conductivity maxima and minima on the microwave power
(tending to saturation in the range of elevated powers),
(c) steep decrease in the maxima and minima magnitude resulting 
in vanishing of ANC
and, hence, in vanishing ZR- and ZC-states with
increasing temperature, and (d) relatively small magnitude of the minima
and maxima in 2DESs with moderate electron mobility
that makes impossible
the attainment of ANC and its consequences.

In this letter, we discuss a scenario for
the appearance of zero-resistance (ZR)
as well as zero-conductance (ZC) states in 2DESs invoking
the concept of ANC associated with photon-assisted impurity
scattering complicated by electron-electron interaction
and photon-assisted acoustic phonon scattering. The proposed scenario
provides plausible explanations of main experimental
facts. 

\section{ANC due to photon-assisted scattering}
The effect of ANC in  a 2DES
system in  magnetic field under microwave irradiation is associated
with the following~\cite{5,12}. The dissipative electron transport in the direction
parallel to  the electric field and perpendicular to the magnetic field
is due to hops of the electron Larmor orbit centers caused by
scattering processes. These hops result in a change in the electron
potential energy $\delta \epsilon = - F\delta\rho$.
Here $F$ is the dc electric force acting on an electron which is determined
by 
the net in-plane
dc electric field including  both the applied and
the Hall components, and $\delta\rho$ is the displacement of 
the electron orbit center.  
If the electron  orbit center
displaces in the direction of the electric force ($\delta\rho > 0$ and  
$\delta \epsilon < 0$),
the electron potential energy decreases.
In equilibrium, the electron orbit center hops in this
direction are dominant, so the dissipative electron current flows in
the direction of the net dc electric field. 
However, in some cases,
the displacements of the electron orbit centers in the direction
opposite to the electric force (with  $\delta\rho < 0$
and, hence, $\delta \epsilon > 0$) can prevail resulting in the 
dissipative current
flowing opposite to the electric field.
Indeed,  if an electron absorbs  a photon
and transfers to a higher LL, a portion of the
absorbed energy $N\hbar \Omega_c$ ($\hbar$ 
is the Planck constant)  
 goes to an increase
of the electron kinetic energy, hence, the change
in the electron potential energy
is
$\delta \epsilon = \hbar(\Omega - N\hbar \Omega_c)$.
If $(\Omega - N\hbar \Omega_c) > 0$, so that $\delta \rho < 0$ (see fig.2a),
 the potential energy of electrons
increases with each act of their scattering.  

\section{Phase of the dissipative conductivity oscillations}

Summarizing the results of previous calculations~\cite{5,12,21} 
(see also~\cite{20,22}),
the variation of the dissipative dc current under the effect
of microwave radiation (photocurrent) can be presented as
\begin{equation}
J_{ph} \propto \sum_{N, M} \Theta_NI_{N}^M\frac{(N\Omega_c - M\Omega)}
{[(\Lambda\Omega_c - M\Omega)^2 + \Gamma^2]^2}
\end{equation}
if $FL \ll \hbar \Gamma$,
\begin{equation}
J_{ph} \propto \sum_{N, M}\Theta_N I_{N}^M(N\Omega_c - M\Omega)\exp
\biggl[- \frac{\hbar^2(N\Omega_c - M\Omega)^2}{2F^2L^2}\biggr]
\end{equation}
when $FL > \hbar \Gamma$ and $L > d_i$, and
\begin{equation}
J_{ph} \propto \sum_{N, M}\Theta_N I_{N}^M(N\Omega_c - M\Omega)\,K_0
\biggl(\frac{|N\Omega_c - M\Omega|d_i}{v_H} \biggr)
\end{equation}
when $FL > \hbar \Gamma$ and $L \ll d_i$ (smooth disorder).
Here $e$ 
the electron charge, $\Gamma$ is the LL broadening, 
 $L$
is the magnetic length, 
$d_i$ is the spacing between 2DES and the donor sheet, $v_H$ is the Hall drift
velocity, $K_0(z)$ is the McDonald function, and
$\Theta_N = 1 - \exp(- N\hbar \Omega_c/T)$,
where $T$ is the electron
temperature. Factor $\Theta_N$ is due to the contribution of scattering processes
with both absorption and emission of microwave photons. 
Coefficients $I_{N}^M$ are determined by the matrix elements
of photon-assisted interaction of electrons with impurities (remote ones
and those
in 2DES) and surface roughness as well as by the amplitude of the ac microwave
electric field $\cal E$ and the electron distribution function. 
As follows from~(1) - (3), the microwave photocurrent
reaches the maxima at $N\Omega_c - M\Omega = \Delta^{(+)}$
and minima at $N\Omega_c - M\Omega = - \Delta^{(-)}$
with $\Delta^{(+)} \simeq  \Delta^{(-)} \sim {\rm max} \{\Gamma, FL/\hbar\}$.
According to~(1) - (3), the net dissipative current approximately coincides with
its dark value at the resonances  $N\Omega_c = M\Omega$.
At $N\Omega_c - M\Omega = - \Delta^{(-)}$
and sufficiently strong microwave radiation (when $|J_{ph}| > J_{dark}$),
the net dissipative dc current $J_D = J_{dark} + J_{ph}$
becomes directed opposite to
the electric field resulting in the instability.
This pattern of the oscillatory behavior of the microwave photocurrent
is in line with qualitative reasonings in the previous
section. It is consistent with the experimental 
results~\cite{6,7,8,14,15,16}.

As shown, the photon-assisted acoustic phonon scattering
processes (see fig.2b) also lead to an oscillatory dependence
of the microwave photoconductivity. However, the phase of these oscillations
is opposite to that in the case of photon-assisted 
impurity scattering~\cite{23,24}. This can add complexity to
the microwave 
photoconductivity oscillations and can even result in their suppression, 
particularly, at elevated temperatures
(see Sec.~6).

\section{Power nonlinearity} 

The dependence of the factor 
 $I_{N}^M$ microwave field  is given by $J_{M}^2(\xi_{\Omega})$,
where $J_{M}(z)$ is the Bessel function and $\xi_{\Omega} \propto {\cal E}$ 
is
proportional to the amplitude of classical oscillations of the electron
orbit center in the microwave field (see,
for example~\cite{19,20}). The terms with $M > 0$
correspond to the transitions with the absorption and emission of
$M$ real microwave photons. 
Thus, the magnitudes of the microwave photoconductivity
maxima and minima ${\rm max}\, \sigma_{ph}$ and $|{\rm min}\, \sigma_{ph}|$,
where $\sigma_{ph} = J_{ph}/E$,
are generally  nonlinear functions
of the microwave power $P \propto |{\cal E}|^2$. 
This is due to the effect of virtual photons absorption and emission
on the electron scattering processes.
At low microwave powers,
 ${\rm max}\, \sigma_{ph} \propto P$ and $|{\rm min}\, \sigma_{ph}| 
\propto P$.
However, when $\xi_{\Omega} \propto {\cal E}$ approaches  $b_M$,
where $b_M$ corresponds to maximum value of  $J_{M}(z)$,
the magnitude of   ${\rm max}\, \sigma_{ph}$ fairly slow increases
with microwave power $P$ 
in line with experimental observations~\cite{6,7} and others. 
This occurs at such powers that
the amplitude  of classical oscillations of the electron
orbit center in the microwave field becomes of the order of $L$.
The pertinent characteristic power $P_{max}$
increases with $\Omega$ approximately
as  $P_{max}\propto \Omega^3$~\cite{19}.
Another consequence of the nonlinear mechanism in question
is that 
at high microwave powers the magnitudes of
maxima and minima corresponding to higher resonances
($\Omega \sim N\Omega_c$ with $N > 1$) are not too small compared
to those near 
the cyclotron resonance($\Omega \sim \Omega_c$).

Slowing down of the increase in  ${\rm max}\, \sigma_{ph}$ 
and $|{\rm min}\, \sigma_{ph}|$ with increasing microwave power
can be also associated with some heating of the 2DES. As shown below,
an increase in the electron temperature leads to broadening
of the LL and, consequently, to smearing of the resonances.

\section{Temperature effects}
As seen from (1), the microwave photoconductivity $\sigma_{ph}$
markedly decreases due to the processes with emission of microwave photons.
This effect becomes essential when the electron temperature
increases from $T < N\hbar\Omega_c \sim \hbar\Omega$ to $T > N\hbar\Omega_c\sim \hbar\Omega$. 
The microwave photoconductivity
maxima and minima also strongly depend on the LL broadening. 
The latter can be rather sensitive to the temperature.
 In particular,
at moderate microwave powers $P$, for the Lorentzian 
shape of the LLs,
one obtains the following temperature dependence:
\begin{equation}\label{eq4}
\frac{{\rm max}\, \sigma_{ph}}{\sigma_{dark}} 
\simeq \frac{|{\rm min}\, \sigma_{ph}|}{\sigma_{dark}}  
\propto P
\frac{(1 - e^{-\hbar\Omega/T})}{\Gamma^3(T)}.
\end{equation}
Here the dark conductivity and photoconductivity
stem from  the scattering processes involving inpurities,
while the value $\sigma_{ph}$ depends on the sharpness
of the resonances and, hence, on the net  LL broadening.
The net LL broadening is determined by
the impurity (and roughness) scattering
processes and by the electron-electron interaction.
The LL broadening due to electron-electron interaction steeply increases
with the electron temperature. 
Taking into account that in the experimental
situations the 2DES Fermi energy
$E_F \gg \hbar\Omega_c$, one can use the 
following temperature dependence~\cite{27}:
$\Gamma_e(T) \propto (T/E_F)^2\ln(\sqrt{E_FRy^*}/T)$,
where $Ry^*$ is the effective Rydberg.
For $f = \Omega/2\pi = 50$~GHz, the factor associated with
the emission of microwave photons in~(4) reduces  approximately by half
with the temperature increasing from 1~K to 3-4~K.
Setting $E_F = 10$~meV,
we find that $\Gamma_e|_{T = 3K}/\Gamma_e|_{T = 1K} \simeq 9$.
Hence, according to (4), 
in a 2DES with high electron mobility (in the absence of magnetic field)
in which $\Gamma$ is determined primarily by the electron-electron
scattering
so that $\Gamma \simeq  \Gamma_e$, 
the span of the dissipative conductivity
oscillations, i.e., the values 
${\rm max}\, \sigma_{ph}$ 
and  $|{\rm min}\, \sigma_{ph}|$ can decrease by
several orders of magnitude
when the temperature increases by a few K.
However, in 2DESs with moderate electron mobility (limited, say, by
residual impurities and interface roughness) 
in which $ \Gamma_i \gtrsim \Gamma_e$, an increase in 
$\Gamma$ with increasing
temperature and, hence, a decrease in   ${\rm max}\, \sigma_{ph}$ 
and  $|{\rm min}\, \sigma_{ph}|$ can be less pronounced
as confirmed by experimental data.

Since photon-assisted acoustic phonon
scattering provides the microwave photoconductivity maxima and minima
at $N\Omega_c \lesssim M\Omega$  and $N\Omega_c \gtrsim M\Omega$, 
respectively,
i.e., approximately at the point where 
photon-assisted impurity scattering yields, on the contrary,
the microwave photoconductivity minima and maxima,
the former mechanism can interfere with the latter one modifying
the oscillations and even effectively suppressing them.
This is possible if  photon-assisted acoustic phonon
scattering
becomes essential with increasing temperature~\cite{23,24}. A marked
intensification of this mechanism occurs when $T \gtrsim \hbar s/L = T_{ac}$,
where $s$ is the speed of sound. In the experimental situations, 
$T_{ac} \simeq 0.5$~K.

\section{Effect of high electron mobility.}
Although the oscillations of microwave photoconductivity
as a function of the cyclotron frequency (i.e., the magnetic field)
were observed in 2DESs with the electron mobility in rather wide range,
sufficiently deep  microwave photoconductivity minima
which can result in ANC were observed only in the samples with 
fairly high electron mobility. In the framework of the model
under consideration, this can be explain as follows.
The relative amplitude of the microwave photoconductivity
oscillations is very sensitive to the LL broadening.
In sufficiently perfect 2DESs with weak scattering of electrons
on residual impurities immediately in the 2DES  and on the
interface roughness, the LL broadening is determined primarily by
the electron-electron interaction. 
Indeed, when the electron sheet concentration $\Sigma_e$
is about the sheet concentration of remote impurities $\Sigma_i$,
the ratio of quantities $\Gamma_i$ and $\Gamma_e$ can be estimated roughly 
as  $\Gamma_i/\Gamma_e \propto (\Sigma_i/\Sigma_e)\exp(- 2d_i/L)$.
The exponential factor in this formula is due a spatial separation
of  electrons and donors which gives rise to an exponential decrease
in the matrix element of electron-impurity interaction.
Hence, at $d_i > L$, one obtains  $\Gamma_i \ll \Gamma_e$. In the experiments
with 2DESs having high electron mobility, $d_i/L \simeq 1.4$, so that
the latter exponential factor  is about of 0.06.
Since the electron-electron scattering processes are effectively
suppressed with decreasing temperature~\cite{27}, 
the microwave maxima and, what is more important,
minima are well pronounced and can surpass the dark conductivity
at low temperatures and when the microwave radiation is strong enough.
This leads to ANC in some ranges of magnetic field when  certain relations
between $\Omega$ and $\Omega_c$ are met. 
In contrast, in the samples with moderate electron
mobility, a significant contribution to the LL broadening
is provided by residual impurities and interface roughness.
This prevents the attainment of a sufficiently large ratio  
$|{\rm min}\, \sigma_{ph}|/\sigma_{dark}$ necessary for ANC. 

\section{Conclusion}
We believe that main experimental facts on ZR- and ZC-states
and related effects can be explained in the framework of the concept
based on ANC caused by photon-assisted impurity scattering of
electrons and affected by electron-electron   and photon-assisted
acoustic phonon interactions.

\end{document}